\documentclass[%
amsmath,amssymb,
aps,
rmp,
]{revtex4-2}

\usepackage{graphicx}
\usepackage{dcolumn}
\usepackage{bm}
\usepackage{hyperref}

\begin{document}
	
	\preprint{APS/123-QED}
	
	\title{Ultra-intense laser pulses and the High Power Laser System 
		at Extreme Light Infrastructure - Nuclear Physics
	}
	\thanks{Lecture notes for Femto Up! School, Strasbourg, France, March 2021}%
	
	\author{Daniel Ursescu}
	\altaffiliation[Also at ]{Physics Department Doctoral School, Bucharest University, Bucharest, Romania}
	\email{daniel.ursescu@eli-np.ro}
	\affiliation{%
		National Institute for Physics and Nuclear Engineering, Extreme Light Infrastructure - Nuclear Physics, Reactorului 30, 077125, Magurele, Ilfov, Romania
	}%

	\date{\today}
	
	\begin{abstract}
		Lasers make possible the production and ultimate control of electromagnetic fields in terms of spectral purity, spatial confinement down to  micrometer scale, duration down to a single cycle in the femtosecond domain or shorter and electromagnetic field strengths - corresponding to the highest intensities achieved by mankind in the lab. Ultra-intense laser facilities are pushing the limits of the achievable pulse intensity, hence the coined term extreme light. They make possible fundamental and applied investigations in physics and material science with emergent societal impact. Extreme Light Infrastructure is the most advanced project dedicated to the production and use of such extreme fields. \\
		The Extreme Light Infrastructure project will be outlined, with emphasis on the extreme light capabilities of the three pillars. The architecture of the first finalized 10PW high power laser system (HPLS) will be highlighted. This dual arm, 10PW each, laser system, at Extreme Light Infrastructure Nuclear Physics (ELI-NP), in Romania, delivers beams in five experimental areas that address research centered on nuclear physics, materials in extreme environments and exotic physics. 
		
%
		
	\end{abstract}
	
	\maketitle
	
	\tableofcontents
	
	\section{\label{sec:level1}Introduction}
	
	The lecture notes here were written for the FEMTO UP School 2021, which took place in on-line form instead of Strasbourg, France due to the COVID19 pandemic.
	
	The present manuscript aims to introduce the reader to the ultra-intense laser pulses and to the machines and technologies used to produce them. To begin, it is essential to place the laser beam in the broader context of fundamental forces in physics and to provide a quick introduction to the concept of laser beam and laser pulse, without going in depth. 
	The specification of the geometry and time distribution in highly intense laser beams is essential for the description of the laser-driven experiments, for the simulations of the laser-matter interaction processes and for the metrology of the laser pulses.
		The Gaussian  laser pulses described here are a simplified model of the reality. However, they are widely used as reference in ultra-intense laser pulses research field because the scientists in the field need to share a unique definition of the object under study to be able to compare results from one team to another.
	The spatial and temporal scales of the ultrashort pulses varies orders of magnitude. As an example, the 10PW pulses from ELI-NP has approximately 50 cm aperture after compression and shall reach 4 $\mu$m in the focal spot while the intial pulse duration from the laser oscillator is 6 fs and reaches 1 ns after the temporal stretching of the pulse, in the final laser amplifier, before compression.
	Also, the real-life pulses often present deviations from the ideal Gaussian pulses taken as a model reference. The unwanted temporal and spatial distortions might generate deleterious hot spots that affect the experimental conditions and harm the laser components, so they are carefully measured and suppressed. But also such deviations are sometimes introduced with the purpose to enhance the experiment results.

	The second part of the presentation will concentrate on the description of the most powerful laser system that operates to date, namely the High Power Laser System (HPLS) at Extreme Light Infrastructure – Nuclear Physics.
	
	The third part will present the international context and technological developments related to the high power laser systems, highlighting concerted efforts of developments worldwide and in particular the Extreme Light Infrastructure pan-European project.

	
	\section{\label{sec:concepts} From fundamental forces to the technology for producing the highest intensities}
	
	\subsection{\label{sec:level2} Laser beam and laser pulse }
	
	Out of the four fundamental interaction types, the electromagnetic one is the easiest to control. This is reflected through the myriad devices and the ubiquitous power plug but also through the applications based on mastering light, from spectroscopic investigations of chemicals to laser eye surgery.
	The detailed behavior in space and time of the electromagnetic field is described through the microscopic Maxwell equations which connect the electric (\textbf{E}(\textbf{r},t)) and magnetic (\textbf{B}(\textbf{r},t))  fields to the electrical current (\textbf{J}(\textbf{r},t))  and charges  ($\rho$(\textbf{r},t)) densities:
	
	\begin{eqnarray}
		\label{eq:Maxwell-total-a}
		\nabla \cdot \mathbf{E}(\mathbf{r},t)& = & \frac{\rho (\mathbf{r},t)}{\epsilon_0}  \\
		\label{eq:Maxwell-total-b}
		\nabla\cdot \mathbf{B}(\mathbf{r},t)& = & 0 \\
		\label{eq:Maxwell-total-c}
		\nabla \times \mathbf{E}(\mathbf{r},t) & = & - \frac{\partial \mathbf{B}(\mathbf{r},t) }{\partial t} \\
		\label{eq:Maxwell-total-d}
		\nabla \times \mathbf{B}(\mathbf{r},t) & = & \mu_0 \mathbf{J}(\mathbf{r},t)+ \mu_0 \epsilon_0 \frac{\partial \mathbf{E}(\mathbf{r},t) }{\partial t}
	\end{eqnarray}
	where $\epsilon_0$ and $\mu_0$ are the electric and magnetic constants of the vacuum, also known as electrical permittivity and magnetic permeability of free space.  

	In the absence of the electric charges and currents, one can deduce the electromagnetic wave equations starting from the Maxwell’s equations. They read:
\begin{eqnarray}
	\label{eq:wave-E}
	\frac{\partial^2 \mathbf{E}(\mathbf{r},t) }{\partial t^2}- c_0^2  \nabla^2 \cdot \mathbf{E}(\mathbf{r},t)& = & 0 \\
	\label{eq:wave-B}
	\frac{\partial^2 \mathbf{B}(\mathbf{r},t) }{\partial t^2}- c_0^2  \nabla^2 \cdot \mathbf{B}(\mathbf{r},t)& = & 0
\end{eqnarray}
where $c_0$ is the speed of light and the relation $c_0^2=\epsilon_0 \mu_0$ holds.
 These equations describe the behavior of arbitrary electromagnetic fields across the entire spectra of frequencies. The equations accept as a solution the sinusoidal function, known as plane wave (pw):
	
\begin{equation}\label{eq:plane-wave}
	\mathbf{E}_{pw}(\mathbf{r}, t,\hat{\mathbf{k}},\nu)= \mathbf{E}_{pw0} \cos(2\pi \nu t–2\pi \frac{ \nu}{c} \hat{\mathbf{k}} \cdot \mathbf{r})
\end{equation}	
	
	with $\nu$ the frequency of the field and $\mathbf{\hat{k}}$ a unit vector. For $\textbf{B}_{pw}$ the form is similar, as the equations \ref{eq:wave-E} and \ref{eq:wave-B} have the same form.  It can be shown that the sum of two particular fields that are solutions for the wave equation is also a solution for the wave equation. As a consequence, the plane wave is widely used in the description of electromagnetic phenomena and there is a widely used mathematical approach to handle it, namely the Fourier transform, which can be understood as a summation of plane waves over the entire range of frequencies and directions: 
		
	\begin{equation}\label{eq:Fourier}
		\mathbf{E}(\mathbf{r}, t)= \int\mathbf{E}_{pw} d\nu d\hat{\mathbf{k}}
	\end{equation}	

	One peculiarity of these wave fields is that they correspond to orthogonal \textbf{E} and \textbf{B} vector fields which oscillate in phase. The propagation direction of the wave is defined through the vector which is locally orthogonal to both electric and magnetic fields. It can be shown form Maxwell’s equations that this direction corresponds to the $\hat{\textbf{k}}$ and the following relation holds:
	\begin{equation}\label{eq:B-E}
			\mathbf{B} =\frac{\mathbf{k}}{\omega} \times 	\mathbf{E}
	\end{equation}

	Before proceeding, it is important to note that one can replace the sine function that we get in the plane wave through a combination of complex exponential functions, thanks to the mathematical identities: 
\begin{equation}
	\cos x =\frac{\mathit{e}^{ i x} + \mathit{e}^{- i x}}{2}=\frac{1}{2}(\mathit{e}^{ i x}+\overline{\mathit{e}^{ i x}})
\end{equation}

where $\overline{\mathit{e}^{ i x}}$ represents the complex conjugate term of $\mathit{e}^{ i x}$. Often, this complex conjugate terms are not explicitly included in calculations, as the math is similar for the complex conjugate part. Then the field of eq. \ref{eq:plane-wave} and \ref{eq:Fourier} are written in the form: 
\begin{equation}
	\label{eq:planeWave-exp}
	\mathbf{E}_{pw}(\mathbf{r},t)  =  
 \mathbf{E}_{pw0} \mathit{e}^{\mathit{i}(\mathbf{k} \cdot \mathbf{r} - \omega t )} 
\end{equation}

and 
\begin{equation}
	\label{eq:expEFourier-exp}
	\mathbf{E}(\mathbf{r},t)  =  
	\int \mathbf{E}_{pw0} \mathit{e}^{\mathit{i}(\mathbf{k} \cdot \mathbf{r} - \omega t )} \mathrm{d}{\mathbf{k}}
\end{equation}

	The plane wave description of the electromagnetic fields has the drawback that it uses functions that are oscillating at infinity with non-negligible amplitudes. Such a single pure plane wave it is impossible to be produced in the lab. Hence, in order to describe electromagnetic fields produced in the lab, that are localized in space, one has to sum up an infinite number of plane waves. This approach allows one to obtain a suppression of the resulted field of the wave beyond a large distance, hence a localized wave packet.
	
	An alternative solution for the wave equations, instead of plane wave solution of equation \ref{eq:planeWave-exp},  can be obtained after several approximations, in the form of the Gaussian beam (see for example chapters 16-19 of \cite{Siegman1986} detailed description of the derivation and mathematical properties). The starting point is the separation of the spatial and temporal parts in the wave equation, assuming a well defined frequency of the light oscillations and a linear polarization for the electric field in order to reduce the equation to a scalar equation for one component:
\begin{equation}
	\label{eq:separation}
	E(\mathbf{r} ,t)=u(\mathbf {r} )T(t)
\end{equation}
	Replacing the result in the wave equation for the electric field \ref{eq:wave-E}, one can separate the the equation in two parts, one depending only on space coordinate \textbf{r} and the other involving only the temporal coordinate.	In this way, one can extract the differential equation for the spatial part of the electric field as: 
		\begin{equation}\label{eq:HElmholtz}
			\nabla ^{2}u(\mathbf{r})+k^{2}u(\mathbf{r}) = 0 ,
		\end{equation}
	while the temporal part is described with the following form:
	\begin{equation}\label{eq:temporal}
		\frac {\mathrm {d} ^{2}T(t)}{\mathrm {d} t^{2}}+c^2 k^{2}T(t)=0.
	\end{equation}
	In addition, one is interested by a solution corresponding to a defined direction in space, direction defined as propagation direction through the $\hat{k}$ unit vector, which is chosen to be along the z axis of the orthogonal reference system without losing the generality of the result.  
	An approximate solution of this spatial part of the electric field u(\textbf{r}) in cylindrical coordinates is then given by the following formula:
	\begin{equation}\label{eq:Gauss-beam}
		u(\mathbf{r}) = {u} (r,z)=u_{0}\,\,{\frac {w_{0}}{w(z)}}\exp \left({\frac {-r^{2}}{w(z)^{2}}}\right)\exp \left[\!-i\left(kz+k{\frac {r^{2}}{2R(z)}}-\psi (z)\right)\!\right], 
	\end{equation}
 where 
\begin{equation}\label{eq:waist}
	w(z)=w_{0}\,{\sqrt {1+{\left({\frac {z}{z_{\mathrm {R} }}}\right)}^{2}}},
\end{equation} 
\begin{equation}\label{eq:waist}
	R(z)=z \left[ {1+{\left({\frac {z}{z_{\mathrm {R} }}}\right)}^{2}}\right],
\end{equation} 
\begin{equation}\label{key}
	\psi (z)=\arctan \left({\frac {z}{z_{\mathrm {R} }}}\right),
\end{equation}

with 
\begin{equation}\label{eq:zR}
	z_{\mathrm {R} }={\frac { \pi w_{0}^{2} \nu}{c}}
\end{equation}
	This solution has the advantage of being localized in space on two directions (x and y axes, by convention) and to form a Gaussian beam when the temporal part is included using the result of equation \ref{eq:temporal}. It is important to note that this is an approximate solution for the Maxwell’s equations, as an approximation of the equation \ref{eq:HElmholtz} is used. Exact analytical solution close to the Gaussian beam form exist (\cite{Barton1989,Salamin2006, Salamin2002}) in the form of an infinite Taylor series, where the lowest order approximation corresponds to the Gaussian beam. 
	
	For the Gaussian laser beams, there are some basic rules concerning the relation between the divergence of the beam and the smallest waist of the beam, w$_0$, 
\begin{equation}\label{eq:waist}
		w_0= \frac{c}{\pi \theta \nu} 
\end{equation}	

	where $\theta$ is the beam half divergence in the far field.
	 One can connect the f$\#$=F/D of a focusing component to the achievable waist w$_0$. D corresponds to collimated Gaussian beam diameter measured at an amplitude equal to  1/e from the maximum amplitude, and F represents the focal distance of the focusing element. Then one can write the following relations, in the approximation of small angle $\theta$ expressed in radian:
	\begin{equation}\label{eq:half-divergence}
		\theta \approx \tan(\theta) =\frac{D}{2F}=\frac{1}{2 f\#}
	\end{equation}

\begin{figure}
	\centering
	\includegraphics[width=0.7\linewidth]{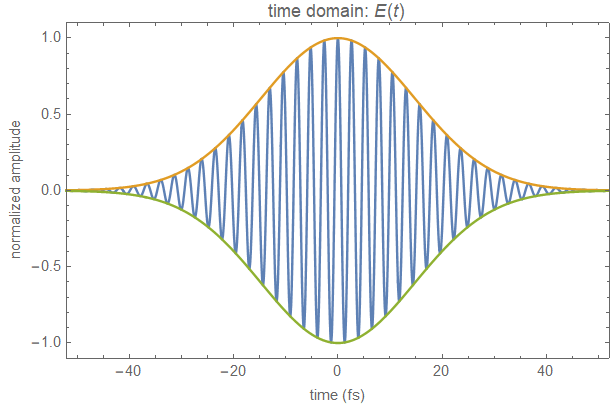}
	\caption{Illustration of a temporal evolution of the field in a Gauss pulse, corresponding to the temporal part of equation \ref{eq:Gauss-pulse}}
	\label{fig:temporal-gauss}
\end{figure}

	In practice there is no infinite laser beam available and, moreover, we are interested in short laser pulses, in the range of few (tens or hundreds of) cycles. The description of such a pulse can be achieved by summing up Gaussian beams with different frequencies.  When the beams are in phase at a well defined point in space and time, the superposition of the beams with various wavelengths will lead to constructive addition of the electric field in a limited region in space and also in time. The result can come close to a Gaussian beam modulated in time with a Gaussian envelope when the amplitudes and phases of the beams are carefully chosen: 
	
	\begin{equation}\label{eq:Gauss-pulse}
			E(\mathbf{r} ,t)=u(\mathbf {r} ) \exp\left[ -\left(\frac{z- c t}{c \tau} \right)^2 \right] \cdot \exp\left(-i  \omega t \right)
	\end{equation}
	 The real part of the temporal terms in the electric field description from equation \ref{eq:Gauss-pulse} is depicted in figure \ref{fig:temporal-gauss}. The   phase information is included in the oscillations of the field amplitude while the Gaussian envelope shows the overall pulse shape.
	 
	Most of the ultrafast laser systems consider the output of the laser in terms of Gaussian pulses. Some of the properties are defined through comparison with an ideal pulse, often considered to be Gaussian in shape, for simplicity.

	\subsection{\label{sec:intensity} Why intensity matters?}

The optical intensity of the electromagnetic field, in vacuum, is then defined by the cycle-averaged modulus of the Poynting vector $\mathbf{S}$:
\begin{equation}
	\label{eq:poynting}
	I = |\mathbf{S} |  =  c^2 \epsilon_0 |\mathbf{E} \times \mathbf{B}| = \frac{1}{2} c \epsilon_0 |\mathbf{E}|^2
\end{equation}
where the equation \ref{eq:B-E} was used. The intensity  has units of Watt/meter$^2$. It corresponds to the electric field energy density multiplied with the speed of light, and corresponds to the transmitted optical power per unit area through an imaginary surface placed perpendicular to the propagation direction of the electromagnetic field \textbf{k}. Optical intensity has the advantage that it can be determined experimentally through measurements of the pulse energy, duration and diameter.

 For a Gaussian beam having the optical power P and beam waist w, one can compute the peak intensity (on the propagation direction of the beam) using the relation:
\begin{equation}\label{eq:intensity-beam}
I =\frac{2 P}{\pi w^2}	
\end{equation}

The peak power of a rectangular pulse in time can then be computed as the ratio between the measured energy of the pulse and its duration (considerded full width at half maximum (FWHM) of the energy envelope). To obtain the peak intensity for a Gaussian temporal pulse, one can then multiply by a correction factor of 0.939.
\begin{equation}\label{eq:gauss-intensity}
	I_G =0.939 I
\end{equation}

\begin{figure}
	\centering
	\includegraphics[width=0.5\linewidth]{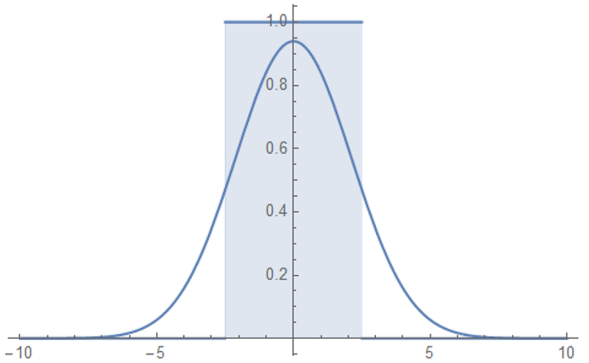}
	\caption{Comparison of a Gauss pulse and a rectangular pulse with the same full width at half maximum and with the same area under the curve, corresponding to the same energy in the pulse.}
	\label{fig:gauss-pulse}
\end{figure}

To illustrate the need of this correction factor,  the energy envelope for a rectangular pulse of width t$_0$ and a gaussian pulse of duration t$_0$ FWHM, having the same integrated energy in time, are represented in fig. \ref{fig:gauss-pulse}. The peak amplitude of the Gaussian envelope reaches only 0.939 from the maximum amplitude of the rectangular pulse.

\begin{figure}
	\centering
	\includegraphics[width=0.7\linewidth]{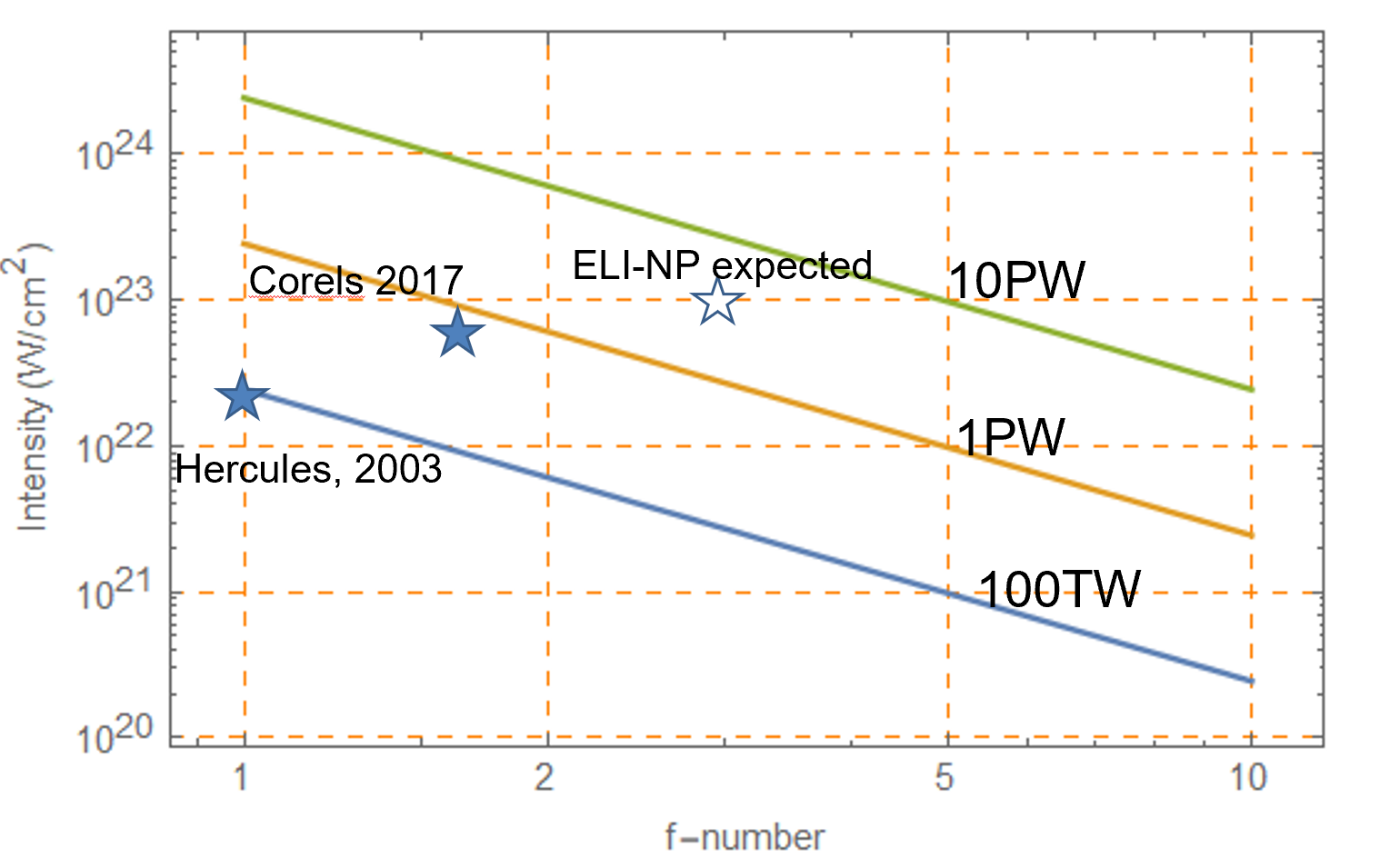}
	\caption{The achievable intensity in reach with 100TW, 1PW and 10PW ideal Gaussian pulses. The stars correspond to measurements at Hercules laser reported  in 2008 and at Corels laser reported in 2019.}
	\label{fig:intensity-fnumber2}
\end{figure}

To illustrate the achievable peak intensities, the figure \ref{fig:intensity-fnumber2} depicts the iso-contours for peak powers of 100TW, 1PW and 10PW  in a plot having the peak intensity on the vertical axis. The stars indicate the previously demonstrated highest intensities \cite{Yanovsky2008,Yoon2019}, reaching $5.5 \cdot 10^{22} \frac{W}{cm^2}$. To be noted that optical components with f$\#$ of the order of unity are extremely costly and difficult to be produced for large aperture beams.

With these details and with equations \ref{eq:poynting} and \ref{eq:B-E} in mind, several practical evaluations of the electric and magnetic fields can be made \cite{Eliezer2002, Mulser2010}.
The electric field associated with the intensity of the pulse is given by the formula:
\begin{equation}\label{eq:electric-numeric}
	E[V/cm]=27.5 \times \sqrt{I[W/cm^2]}
\end{equation}
while the magnetic field can be described with a similar relation, depending only on intensity:
\begin{equation}\label{eq:magnetic-numeric}
		B[Gauss]=9.2 \times 10^{-2}\sqrt{I[W/cm^2]}
\end{equation}

The magnetic field produced with intensities of the order  $10^{22} \frac{W}{cm^2}$  \cite{Yanovsky2008,Yoon2019} are the largest ever produced in controlled manner. The associated magnetic field of such laser pulses reaches 9.2GGauss or 920kT. In comparison, the magnetic field generated indoor reaches 1.2kT \cite{Nakamura2018}  and produced in plasma with the help of a laser of almost $10^{20} \frac{W}{cm^2}$ intensity was 34kT \cite{Tatarakis2002}. Methods complement each other and the fields exist over different time scales.
One can also compute the associated light pressure \cite{Eliezer2002} for a laser pulse intensity as: 
\begin{equation}\label{eq:pressure-numeric}
	p[bar]=3.3 \times 10^{-2}I[W/cm^2] (1+R)
\end{equation}

Where R is the surface reflectivity. A pressure similar to the one in diamond anvil cells can be achieved with intensities of the order of $10^{16} \frac{W}{cm^2}$, an intensity several orders of magnitude below the achievable intensities in modern ultra-intense lasers. To be noted that, unlike in the case of electric or magnetic field in the laser pulse, the light pressure scales linear with the intensity.

From the above comparisons, it is obvious that achieved magnetic fields and pressures are above what is currently achievable in controlled experimental conditions. When comparing electrical fields with the ones that can be observed in the laboratory, this is no more the case. 
The electric fields that can be obtained with ultra-intense laser pulses are smaller than electric fields that are associated with the electrostatic (Coulomb) forces in the inner shells of the atoms. This is due to the fact that the Coulomb force has a proportionality with 1/r$^2$ hence it is singular (infinite) in the origin. 
Regions with much higher fields exists close to the nucleus of the atoms. For example, the Bohr radius associated with H-like atoms is provided by the formula: 
Bohr radius 
\begin{equation}\label{eq:Bohr-radius}
	r_{Bn}= 0.529 \AA \cdot \frac{n^2}{Z}
\end{equation}

 where Z is the number of protons in nucleus and n the energy quantum number.  Hence the electric field at that orbiting distance reaches 
 \begin{equation}\label{eq:Bohr-field}
 	E_{rBn}=\frac{1}{4\pi \epsilon_0}  \frac{Z e^2}{r_{Bn}^2}\sim Z^3
 \end{equation}

However, the electric field associated with ultra-intense laser pulses is strong enough to suppress the Coulomb forces for the outer bound electrons in atoms, for up to 10 electrons, inducing multiple direct field ionization. This process can take place in the volume of the focal spot region which is huge when compared with the volume of an atom. The volume of the focal region where the intensity is significant to induce multiple direct ionization, exceeding several cubic micrometers, can contain more than 10$^9$ atoms.


Another practical quantization of the electric field strength is to compare the gained energy of an electron in the electric field with the rest energy of the electron m$_e$c$^2$. This provides the so-called normalized vector potential a. This non-dimensional parameter is one when the energy is similar to the rest mass hence a relativistic treatment of the electron dynamics is mandatory. 
The a parameter depends not only on the intensity of the pulse, but also on the wavelength of the field, according to the formula:

For intensities larger than $10^{18} \frac{W}{cm^2}$  and wavelength of $\mu$m, this becomes a$\approx$1. At such intensities, the electron dynamics needs relativistic description.

Of course, these evaluations correspond to transient electromagnetic fields and pressures over extremely short time scales, of the order of the laser pulse duration, typically in the femtosecond range.
\subsection{\label{subsec:contrast} Temporal contrast}
The real shape of the laser pulses is not perfect Gaussian in time. Often, the pulse is accompanied by pre-pulses, post-pulses or a pedestal of light (shortly background light) at an intensity level several orders of magnitude lower than the peak intensity. Even in such cases, the background light can produce significant damage to the target, often transforming it to a plasma before the main pulse arrives.

The three typical sources of background light are the longitudinal amplified spontaneous emission in the amplifiers, multiple reflections on the surfaces of optical components that operate in reflection and also the light scattering from the diffraction gratings used in optical stretchers or compressors \cite{Hooker2011}.

The methods that are implemented to enhance the temporal contrast of the pulses include the use of the polarization rotation in a nonlinear crystal with third order non-linearity \cite{Jullien2005, Jullien2006}. The crystal is placed between two crossed polarizers, hence the name of cross polarized wave (XPW) generation. Only the light field generated through the non-linear process with the appropriate polarization is further amplified. The XPW is the temporal domain analogue to the optical spatial filter.

A complementary method for temporal contrast enhancement is related to the use in the laser amplifiers of optical parametric amplification process instead of using laser active media with population inversion and storage of energy \cite{Dubietis2006}. The seed pulse is amplified through energy transfer from a laser pump pulse, mediated by a non-linear crystal. The specific advantage in this case is given by the fact that there is no amplified spontaneous emission outside the time interval defined by the pump pulse duration. 

A third approach to enhance the temporal contrast of the pulse in the few picoseconds interval before the arrival of a main laser pulse is with the help of plasma mirror \cite{Dromey2004, Thaury2007}. This can be understood as an ultrafast reflectvity switch. The intense laser pulses reaching an optical surface of the transparent material can generate, when high intensities (typically above 10$^{14}$ W/cm$^2$), a plasma that behaves as a mirror. All the prepulses at lower intensities pass through the transparent material while the high intensity part of the pulse is reflected and used for experiments. The temporal contrast of the laser pulse can be enhanced about two orders of magnitude in this way. 

The most common method to measure the temporal contrast is based on non-collinear third order autocorrelation of the pulse in a non-linear crystal \cite{Luan1993, Stuart2016}. The pulse is split in two pulses and one of the two replica is frequency doubled. Then a non-collinear autocorrelation curve is measured between the second harmonic replica and the other pulse, through delay scanning. The signal is detected usually using a photo-multiplier. Several companies are selling devices that can cover more than 11 orders of magnitude in dynamic range to qualify the picosecond temporal contrast of ultrashort pulses. Also, new methods to measure the temporal contrast in single shot manner are actively investigated \cite{Oksenhendler2017, Wang2014, Wang2019}.
	
	\subsection{\label{sec:CPA} Chirped Pulse Amplification (CPA) architecture}
	One specific aspect in most high power laser system is the need of significant energy in the laser pulse. Handling the amplification of pulse energy in laser systems involves the use of optical components that shall not be affected by the laser pulse. Common issues to be considered are surface damage of optical components and non-linear effects such as self phase modulation. 
	The laser induced damage takes place for fluences in the range from tens of mJ/cm$^2$ to few J/cm$^2$, depending on parameters such as pulse duration, pulse repetition rate, and of the optical component preparation. For laser pulses below 100fs pulse duration, such components are able to handle typically between 100mJ/cm$^2$ and 300mJ/cm$^2$, corresponding to tens of square centimeters for optical components to be used with pulse energies of the order of 10J. 
	The self phase modulation is a process associated with the non-linear behavior of the refractive index of a material, also known as optical Kerr effect. For high intensity laser, the refractive index of most materials can be described with the formula:
	\begin{equation}\label{eq:n2}
		n=n_0+n_2 I
	\end{equation}
	where n$_0$ is the linear part of the refractive index of the material and n$_2$ is the non-linear part of the refractive index. Typical values for the refractive index \cite{Kabacinski2019} correspond to 1-16 $\cdot 10^{-20}$ m/W. When the pulse propagates through a material, the associated optical path for a ray can be written as:
	\begin{equation}\label{eq:opath}
		OP=n \delta = n_0 \delta +n_2 I \delta = OP_l+OP_{nl}	,
	\end{equation}

	where the first term corresponds to the usual, linear part of the refractive index while the second part can be associated with the non-linear part of the refractive index. For the spatial regions of the laser pulse with high intensity, the non-linear part of the optical path can become significant but it remains negligible at the side of the beam, where the intensity is low. As a consequence, the material starts to behave as a gradient refractive index lens that focuses the pulse. Through the focusing of the laser pulse, the intensity further increases and the non-linear part of the optical path is further increasing locally, inducing spectral modulation and filamentation of the beam or even to damage of the material. This is a good example of positive feedback loop which is extremely difficult to control. To avoid the self phase modulation effect, the cumulated break-up integral (B-integral), defined as the phase associated to the non-linear part of the optical path:
\begin{equation}\label{eq:b-integral}
		B=2 \pi \frac{OP_{nl}}{\lambda}  =\frac{2 \pi}{\lambda} \int n_2 I dz 
\end{equation}
	has to give values lower than 1 radian, corresponding to wavefront distortions of $\lambda$/6.
	
	Trying to directly amplify ultrashort pulses, with duration below 100fs, to energies of the order of 1J, in solid or liquid active media with reasonable sizes, quickly increases the B-integral to values which are above this value. As a consequence, direct amplification of ultrashort pulses to high energies would be extremely difficult.
	To overcome this problem, Donna Strickland and Gerard Mourou \cite{Strickland1985} used a method to reduce the intensity (hence B-integral) through stretching in time the pulses to be amplified.
	 After amplification the pulses can be compressed back to short duration using a so called optical compressor. This approach, known  as Chirped Pulse Amplification (CPA), allows now to extend the pulse duration up to 100 000 times, amplify and re-compress them close to the Fourier limit pulse duration. 
	 The technique was rewarded with the Nobel prize for Physics in 2018 for making possible intensities in excess of $10^{22} \frac{W}{cm^2}$ when pulses from CPA laser systems are focused and also for their use in applications, including eye surgery. In the following section, the most powerful (10PW) functional laser system reported in a scientific journal \cite{Lureau2020}, High Power Laser System (HPLS) at ELI-NP facility, will be presented, followed in section \ref{sec:ELI} by the review of the implementation status of the CPA laser systems at the Extreme Light Infrastructure.

		\section{\label{sec:HPLS} ELI-NP HPLS laser system}

This section intends to present a way to read the description of ultra-intense laser systems, such as HPLS, keeping intensity and temporal contrast in mind. It uses the images from the reference \cite{Lureau2020} which presents a more detailed technical description of the HPLS system and its performance. The HPLS laser system is representative for the state of the art in the high power laser systems in 2020, as pointed out in section \ref{sec:ELI}. It was built to specifications by Thales company and installed at the National Institute for Physics and Nuclear Engineering, placed at the outskirts of Bucharest, Romania.

\begin{figure}
	\centering
	\includegraphics[width=0.7\linewidth]{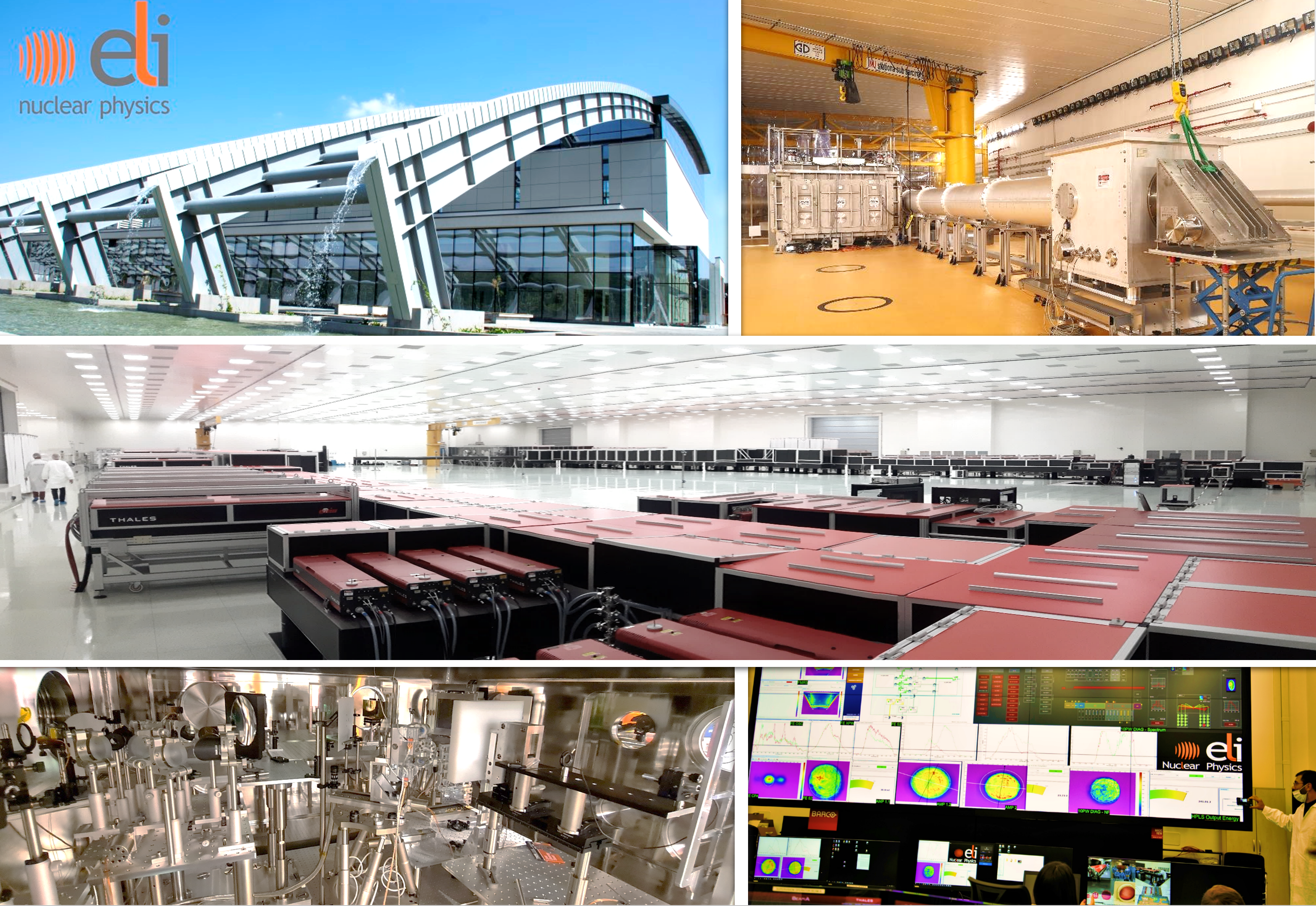}
	\caption{Upper row, left: ELI-NP special building, that hosts the laser system and the experimental areas. Upper row, right: E1 experimental chamber for 2x10 PW pulses and the beam transport for the long focal distance mirror for 10 PW pulses at E6 experimental area. Middle row: laser room, hosting the two laser amplification arms; Lower row, left: inside the 1 PW experimental chamber in E5 experimental area; lower row, right: the large screen in laser control room showing synthetic information on the HPLS status.}
	\label{fig:5images-eli-np-v5}
\end{figure}

The laser is hosted in a clean room  with an area of 2800 m$^2$, as shown in the central image of fig. \ref{fig:5images-eli-np-v5}. The building that hosts the laser, the control room, the experimental area that host the beam transport and the inside of the 1PW experimental chamber in preparation for ion acceleration experiments are also presented in fig. \ref{fig:5images-eli-np-v5}. 

\subsection{HPLS architecture}
\begin{figure}
	\centering
	\includegraphics[width=0.9\linewidth]{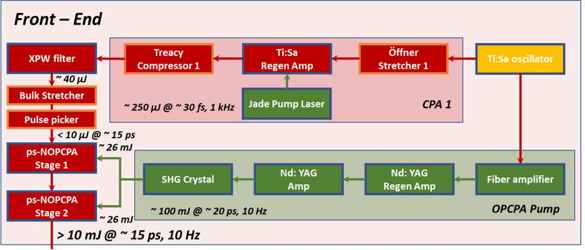}
	\caption{Front-end of HPLS. Copyright CC BY \cite{Lureau2020}}
	\label{fig:frontend}
\end{figure}
The ultra-intense laser systems can be split in three main parts, front-end, amplifiers and compression. The front-end includes a laser oscillator, optical stretcher and initial amplification to an energy level that can be achieved at high repetition rate, in this case at 10Hz, with reasonable cost. The front end also includes  most of the non-linear shaping and control applied to the laser pulses.

 The HPLS front-end is depicted in fig. \ref{fig:frontend}. Here, the front-end starts with a Ti:Sa oscillator that delivers 6 fs pulses at 800nm central wavelength with a bandwidth that is extended to 1064 nm. The oscillator pulse is optically stretched,  amplified in a cavity-based amplifier that runs at 1 KHz repetition rate thanks to the diode pumped Jade laser, and subsequently compressed close to the Fourier limit. The high repetition rate front-ends allow to close fast feedback loops such as the ones that stabilize beam pointing or carrier-envelope phase drift. Once reasonable level of energy is achieved in this way, the next critical aspect, the temporal contrast, has to be tackled. This is made using the cross polarized wave filter (XPW) principle, as described in section \ref{subsec:contrast}. This non-linear filter of the pulse enhances the temporal contrast, broadens and smooths the spectrum but also significantly reduces the useful energy of the pulse. The seed for the XPW filter has to be at best compression, this is why the Treacy compressor is present after the regenerative amplifier (see fig. \ref{fig:frontend}).

 In order to recover the pulse's energy without affecting the temporal contrast, two non-colinear optical parametric chirped pulse  amplifiers (OPCPA) are then implemented. The energy increases three orders of magnitude, from 10$\mu$J to 10mJ. The temporal window of the amplification is defined by the 20 ps pulse duration of the OPCPA pump pulses. These are produced starting from the same oscillator, with the major benefit of optical synchronization with intrinsic low temporal  jitter, in the femtoseconds domain, in contrast to the electronic synchronized  systems where the jitter easily transcends into the picoseconds domain. A narrow spectral part of the oscillator pulses around the 1064nm wavelength is selected using a fiber Bragg grating and then amplified using established technologies including fiber and Nd:YAG crystals as active media. The amplified pulses at 1064nm wavelength are frequency doubled and then used for pumping the OPCPA amplifiers. The OPCPA pump laser is relatively complicated and the low jitter synchronization of the seed and the pump adds sophistication. This is compensated by the added value of the OPCPA stage, namely the temporal contrast enhancement, up to three orders of magnitude outside the temporal amplification window defined by the pump pulse.
 
\begin{figure}[b]
 	\centering
 	\includegraphics[width=0.99\linewidth]{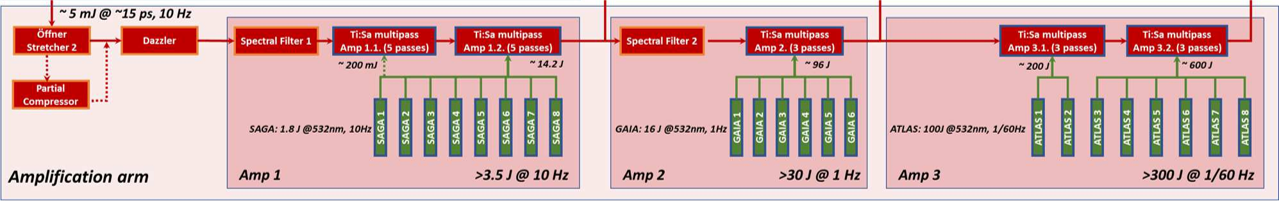}
 	\caption{Amplifiers Amp1, Amp2 and Amp3 of HPLS. Copyright CC BY \cite{Lureau2020}}
 	\label{fig:amplifiers}
 \end{figure}

 At this stage, the 10mJ output of the front-end is divided in two equal energy pulses and sent in two similar, parallel amplification chains, as depicted in fig. \ref{fig:amplifiers}. As discussed in  \ref{sec:CPA}, the pulse has to be stretched significantly in order to avoid self phase modulation and damage of the optical components in the amplification chain. An optical stretcher that brings the pulse to 1ns duration is appropriate for the amplification up to 350J but it is not necessary for the lower amplification levels. For going up to 40J pulses, 600ps would be enough and the major advantage would be to keep the corresponding 100TW and 1PW  compressors more compact. This is why there is a partial compressor implemented after the 1ns stretcher, that reduces the pulse duration to 600ps when the specific 1PW or 100TW compressors are used. 
 
 Also at this place, an acusto-optic programmable dispersion filter is implemented,  as the beam size is still small. It has the role of controlling the spectral phase, hence it helps to fine tune the pulses in order to achieve best compression at the output of the HPLS, corresponding to the highest intensity of the pulse. 
 
 Amplification is realized in Ti:Sapphire crystals using  pump lasers that operate at 10Hz repetition rate with moderate energy of 1.8J at 532 nm wavelength (Amp1 in fig. \ref{fig:amplifiers}). Once the pulse reaches 4J, much more energy is needed for pumping the crystal and this presents the challenge of producing it at 1Hz repetition rate. Here the six pump lasers are delivering up to 96J total energy in the crystal of the second amplifier (Amp2 in fig. \ref{fig:amplifiers}).

The last stage of amplification is based on two more large Ti:sapphire crystals, pumped with a total of eight laser systems delivering each up to 100J per pulse at 1 shot per minute repetition rate. To be noted that all pump lasers for the last amplifiers are using Nd:glass and are flash-lamp pumped. Their thermal management is the limiting factor that determines the repetition rate for the operation of the amplifiers. Solutions similar to those investigated at L2 line in ELI-Beamlines, at HF-2PW in ELI-ALPS or at the PW-class laser system from the group of Jorge Rocca \cite{Wang2017} would make possible to significantly increase the repetition rate of the amplifiers. 

One important aspect in the amplification section of the CPA laser systems is related to the preservation of the  bandwidth for the pulse after amplification. This in a non-trivial task, as the gain shape and  limited bandwidth tends to amplify a narrower portion of the input pulse spectrum and might also shift the  central wavelength. To mitigate this gain-narrowing process, spectral filtering in front of the amplifiers is helpful. These filters are flat mirrors with tailored spectral reflectivity curves, reduced  to 60\% at around 825nm wavelength, as proposed in \cite{Giambruno2011}.

\begin{figure}
	\centering
	\includegraphics[width=0.9\linewidth]{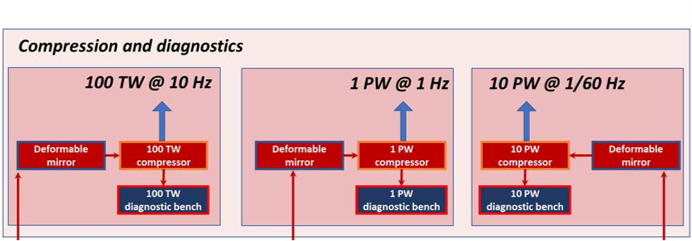}
	\caption{The deformable mirrors, the compressors and the diagnostic benches associated with each of the three main  amplifier systems of HPLS. Copyright CC BY \cite{Lureau2020}}
	\label{fig:compressor}
\end{figure}
The  energy from each amplifier stage is then directed to a dedicated combination of adaptive mirror, optical compressor and diagnostics bench, as depicted in fig. \ref{fig:compressor}. The adaptive mirror is the spatial equivalent of the acusto-optic programmable dispersion filter, namely it adjusts the wavefront of the laser pulse in order to obtain the smallest focal spot from the laser pulse, hence the highest intensity.

The compressors are using reflective diffraction gratings, in order to avoid the self phase modulation of the pulses in bulk optical materials for the compressed pulses. They are also placed in vacuum, in order to avoid the self phase modulation of the pulses in air (one meter of propagation in air is similar to 1 mm propagation in glass, from the non-linear refractive index perspective). The 2.4 J energy from the Amp1 in conjunction with the pulse duration of about 24 fs corresponds to a peak power of 100TW at 10Hz repetition rate for each arm. For the Amp2, the achieved peak power exceeds 1PW at 1Hz repetition rate whiole the 230J and 23 fs pulses of Amp3 correspond to 10Pw peak power at one shot per minute repetition rate.  The damage threshold of the gratings is typically around 100mJ/cm$^2$. This value, in combination with the intended output energy determines the beam size in and after the compressor. In the HPLS case, the 100TW output has 55mm, the one from 1PW has 200mm and the one from the 10PW pulses is about 550mm full aperture beam diameter.

The diagnosis benches implemented after compressors contain the essential tools for qualification of the pulses. They include spatial characterization devices in the form of video cameras for measuring the collimated beam profile, the shape of the focused beam and also the wavefront of the pulse. On the temporal part, there is a self-referenced spectral interferometry device for the spectral phase measurement of the pulse, a spectrometer, a single shot second order autocorrelator for the pulse duration characterization and also a spetrometer. The energy is also monitored with a calibrated photodiode. All these three sets of devices, for spatial, temporal and energy characterization provide the information about the achievable intensity at the specific output of the laser. In addition to these devices, a third order non-collinear autocorrelator with more than 11 orders of magnitude dynamic range is used to qualify the temporal contrast of the pulses.

\subsection{HPLS outputs qualification}
\begin{figure} [htb]
	\centering
	\includegraphics[width=0.9\linewidth]{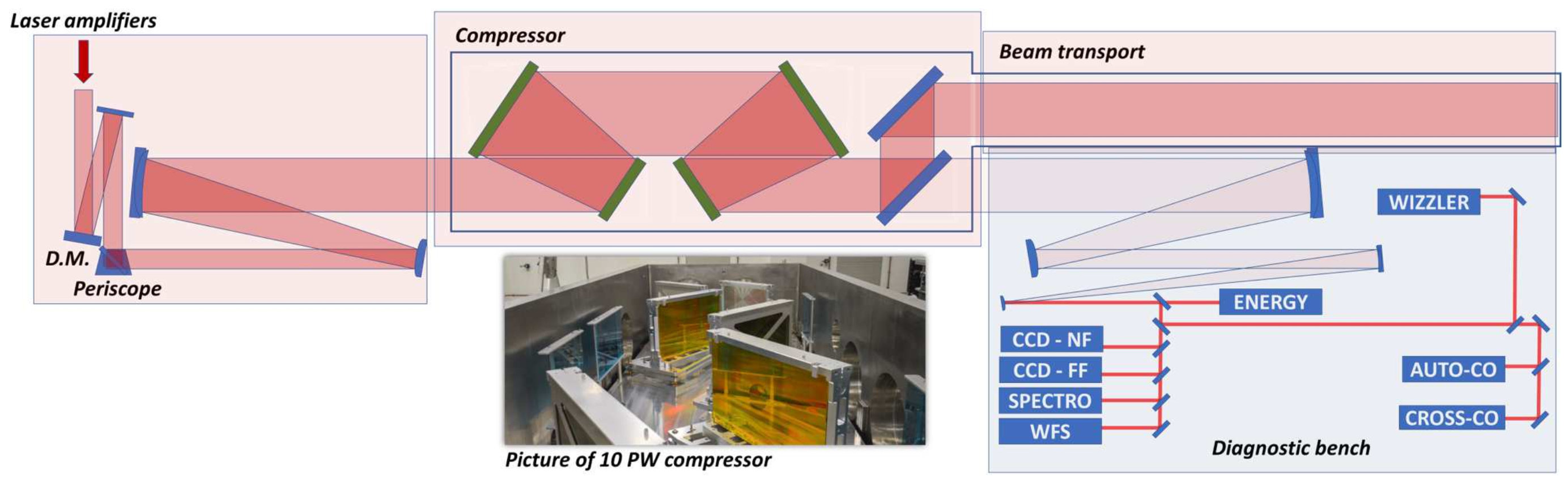}
	\caption[SEt-up for HPLS output characterization at 10PW.]{Set-up for HPLS output characterization at 10PW. The HPLS 10 PW compressor and diagnostics diagram; the inset is a picture of one of the two ELI-NP 10 PW compressors using the meter size gratings. D.M., deformable mirror; WFS, wavefront sensor; CCD-NF, near-field CCD; CCD-FF, far-field CCD; AUTO-CO, single-shot autocorrelator; CROSS-CO, third-order cross-correlator.  Copyright CC BY \cite{Lureau2020}}
	\label{fig:diag-measure}
\end{figure}
The pulses after compressors are sampled at full aperture through a leaky mirror and then they are extracted from vacuum back to air through a window. Then they are de-magnified using a set of telescopes to reach few mm in beam diameter, as suitable for the diagnosis devices used.  The laser pulse sampling method  for metrology purposes might introduce noise and distortions in the measurements. It is the subject of careful calibration and distortions compensation for all specific parameters that are relevant for the determination of the focused pulses intensity.

The measurement of the beam profile and of the wavefront of the pulses is camera-based in all high power laser systems characterization benches. The size of the beam is exceeding the size of the largest chip-sets for such video cameras. This is why the de-magnification of the beam is mandatory. In this respect, the telescopes used for reducing the beam diameter on the sensor are part of the beam profile and wavefront measurement system. The distortions are coming from the alignment precision, from the surface and spectral flatness quality of the optical components used. The most challenging part is the wavefront distortions introduced by the sampling system. In the HPLS case, these distortions were separately measured using an auxiliary laser in a double pass configuration which was sent through a high quality, small aperture beam splitter placed in front of the wavefront sensor through the telescopes and back reflected with a high flatness mirror back on the same optical path to the wavefront sensor. The recorded distortions were below a tenth of a wavelength in our case and they are now subtracted from the raw wavefront data in order to provide accurate information on the pulse wavefront after compressor. 

On the temporal and spectral characterization side, a similar situation appears. The pulses at the exit of the compressors pass through a specially designed flat spectral transmission leaky mirror and  through the compressor window. The total thickness of the traversed  glass is determined by the size of the beam: it has to be al least a tenth of the beam size, in order to reach the specified flatness during the polishing, coating and installation of the mirror. The window shall not collapse under the 1 bar pressure difference exerted on it by the ambient air, when the compressor is under high vacuum. At the beam diameters mentioned above, the dispersion introduced by the glass is significant: for the 55 mm at 100TW output it corresponds to more than 35mm of glass and fourier limited pulses temporally stretched in excess of 160fs, while for 1PW and 10PW pulses the dispersion contribution corresponds temporal stretched pulses longer than 500fs. However, through careful design of the system, the acusto-optic parametric dispersive filter implemented in the amplification chain at the entrance in Amp1 (see fig. \ref{fig:amplifiers}) can compensate for the mentioned spectral phase distortions corresponding to the dispersion of the glass. 

Finally, the energy measurements provide two further challenges. First of all, the aperture of the beam reaches dimensions for which energy measurements devices in femtosecond domain at tens and hundreds of Joule range are prohibitively expensive. Here it is customary to characterize the attenuated beam using a reduced beam diameter before sending the pulses in the optical compressor in vacuum, using available energimeters on the market. Further, the transmission of the compressor is measured or compared with the diffraction efficiencies of the gratings compressors used. The typical values for the diffraction efficiency for one grating range from 90\% to 94\% for a high quality grating. As the optical compressors use four diffraction gratings, their overall transmission lies in the range 65\%-78\% and degrades easily in time if the gratings are contaminated with thin layers of residuals in vacuum. Second challenge is related to the suppression of the full energy, ultrashort  pulses that are measured. A suitable beam-dump has to be built in order to absorb the energy of the pulse after successfully passing through the compressor. Hence, it has to be placed in vacuum and it shall have large aperture, good absorption and good thermal management in the case of high repetition rate lasers. It shall also be positioned in such a way that no back reflection from the beam-dump finds the way back to the amplification chain. Such an unwanted event might generate a focused pulse that destroys optical components in the amplification chain up to the front-end.

 The attainable peak power of HPLS at full amplification was qualified \cite{Lureau2020} using the set-up  depicted in figure \ref{fig:diag-measure}. The spatial properties of the laser pulses, the collimated beam profile, the focal spot and the  wavefront were measured after the demagnification with the telescopes. The wavefront was corrected with the help of the deformable mirror DM placed before the periscope and beam expander in front of the optical compressor. The wavefront calibration as performed as detailed above. On the temporal part, the pulse was measured using the self-referenced spectral interferometry \cite{Oksenhendler2010}. The spectral phase can be controlled with the acusto-optic programmable dispersion filter to achieve best compression  on diagnostics bench, of 22.7fs. The energy  measurement was calibrated. The estimated energy after compression, of 242.6J,  is obtained after subtracting the losses of the compressor, calculated from the diffraction efficiencies of the gratings to be 74.2\%. When taking into account the temporal shape of the pulse as measured on the diagnostics bench, the pulse is not Gauss-shaped; the correction factor in formula \ref{eq:gauss-intensity} is numerically computed to take the value  of 0.87. Hence the achievable power demonstrated at that moment was 9.34PW.

\subsection{Beam transport to experimental areas}
		
	The laser pulses qualification at the output of the laser system is essential for the performance of the experiments, but it is not sufficient. The beam has to be transported to the experimental area in vacuum, avoiding self phase modulations. There, the pulse is focused often using off-axis parabolic mirrors to achieve highest intensity for experiments.
		
	The transportation of such large aperture beams, involving meter-sized mirrors in vacuum is challenging. Several aspects have to be properly taken into account, as detailed in \cite{Ursescu2016}. The challenges are related to the pointing stability, temporal contrast, polarization control, wavefront control, spectral phase control for the pulses and also back reflection isolation. 
	
	The laser beam transport system (LBTS) implemented at ELI-NP was built with simplicity in mind. After exiting the laser room, the LBTS uses only one or two flat mirrors in order to deliver the pulses to the focusing mirror.
	
	The strategy for alignment of the laser pulses to be delivered with micrometer accuracy on the target in focus includes the use of auxiliary guiding laser beams, monitoring of position and direction of the pulses at each mirror, automation and remote control of the mirrors using high precision, high load, vacuum actuators.
	
	The vacuum system is modular, allowing the independent access and maintenance for the two 10 PW pulses. The final vacuum sections of the LBTS are interfaced with full aperture gate valves with the vacuum interaction chambers in the E1, E6 and E7 experimental areas. The vacuum chambers for E1 and E6, simillar in dimensions (3 m length,  4 m width and 2 m in height) are already installed and connected to the LBTS, with the vacuum level reached below 10$^{-6}$ mbar.

	\section{\label{sec:ELI} Extreme Light Infrastructure }
Ultra-intense pulse laser facilities blossomed in the contemporary scientific and technological scenery. They represent state-of-the-art scientific tools with outstanding properties that allow to investigate matter and applications at time scales not accessible otherwise until now. As presented in the previous sections, the magnetic and pressure fields produced in the form of laser pulses correspond to the largest ones created in controlled manner by the human kind. Towards the end of the second decade of the 21st century, more than 130 operational laser systems demonstrating peak power above 10TW or intensities exceeding 10$^{19}$W/cm$^{2}$ were identified by the ICUIL (
\href{https://www.easymapmaker.com/map/ICUIL\_World\_Map\_v3}{\textbf{https://www.easymapmaker.com/map/ICUIL\_World\_Map\_v3}})
. An extended review of the on-going developments towards ExaWatt laser facilities can be found in: \cite{Danson2019}.
In Europe, a strong network of advanced laser laboratories known as LASERLAB Europe was funded over two decades, with outstanding results. Currently, the network of laboratories include 24 laser facilities that provide transnational access for experiments based on scientific merit of the proposal.

With the strong support of LASERLAB, European Union included on the European Strategy Forum on Research Infrastructures (ESFRI) Roadmap an ambitious project of a pan-European distributed research facility dedicated to the most novative light sources and their applications, known as Extreme Light Infrastructure (ELI). The facility is advancing complementary laser technologies in three countries, namely Czech Republic, Hungary and Romania. A fourth pillar shall host a laser system capable to exceed the 10 PW actual technology barrier and to access in the 100 PW region and beyond.

The proposal was backed by an extended study on the relevance of extreme light for the future of the knowledge-based European society and for the mainkind in the form of a White Book \cite{ELI-Whitebook}. We refer the reader to the available public text for an overall image of the fundamental and applied research t obe addressed, ranging from studies of the quantum vacuum structure and particle accelerators to life sciences and management of nuclear waste.

 In the following, short description of the most advanced laser sources at each ELI facility is provided, while secondary sources from THz region to extreme ultraviolet attosecond pulses and gamma rays are beyond the scope of this lecture.

\subsection{ELI-Beamlines}
The pillar in the Czech Republic, known as ELI-Beamlines hosts four major laser systems L1-L4. 	
The L1 ALLEGRA laser system \cite{Batysta2014,Batysta2016,Bakule2020} is a Ti:Sa oscillator seeded OPCPA system pumped at 515nm. It is operated at 1kHz and delivers pulses of about 15 fs after optical compression from 3 ps, using more than 30 bounces on chirped mirrors. It already reaches 30mJ, one third of the design value for the pulse energy to be achieved after amplification in the sixth and seventh amplifiers (100mJ), corresponding to the average power of 30W. The temporal contrast of 10$^{-9}$:1 at 2 ps from the main peak is another unique feature of the laser system.  It already produced UV light at 20nm wavelength for the first experiments.

The L2 Amos laser system is a 100TW laser system designed to operate at more than 20Hz, delivering 2J pulses in 20fs. It uses a 10J pump laser head developed by Rutherford Appleton Lab, U.K. When it will be ready in 2021 or 2022, it will serve as the driver for generating the needed electron bunches for a free electron laser \cite{Wills2020}.

The L3 HAPLS laser system is planned to be used for laser-plasma diagnosis \cite{Tryus2020}, for proton acceleration \cite{Margarone2018} and for electron acceleration \cite{Levato2018}.
The system is using amplification in  Ti:Sa crystals at 800nm wavelength, coupled with diode pumping technology. It already demonstrated a repetition rate of 3.3Hz with an energy of  13.3J and pulse duration of 27.3fs, corresponding to 0.49PW peak power and an average power of about 44W. The nominal design parameters are pulse energy of 30J at a repetiton rate of 10Hz with a pulse duration of less than 30fs.

The L4 ATON laser system  is a kJ-class laser system with two planned outputs, one for long pulses, in nanosecond domain (L4n) \cite{Jourdain2020}, and one intended to deliver pulses of 150fs after compression in vacuum (L4f)  \cite{Cheriaux2019}.
Unlike the other systems at ELI-Beamlines, L4 operates using amplification in Nd-doped glass at 1053 nm central wavelength, at a targeted repetition rate of 1 shot per minute. The beam shape is a 20$^{th}$ order super-Gauss-ian square with 32cm side length. The energy for the long pulses reaches 1.9 J and the temporal shape of the pulse can be adjusted in the range from 100 ps to 10ns with 150ps precision. 

The amplifiers for L4f output demonstrated already up to 1.5kJ energy, and enough bandwidth to support 150fs Fourier limited pulses at one shot every 7 minutes repetition rate. The compression of such pulses remains challenging and the targeted transmission efficiency for it is estimated to be in the 60\% range, corresponding to 6PW output peak power, when the gratings-based optical compressor will be built.


\subsection{ELI-ALPS}
In ELI-ALPS, the core developments are addressing technologies for accessing time scales below 1 femtosecond, namely the attosecond regime, at high repetiton rates. In this way, the attosecond physics can be successfuly addressed \cite{Krausz2009, Ciappina2017}. The sources developed here shall provide electromagnetic pulses in the spectral domain from THz to x-rays.

The 100KHz laser systems developed here, HR1  and HR2 \cite{Haedrich2016, Nagy2019}, share the same approach for the amplification, namely the use of coherent combination from 8 diode-pumped amplifiers, delivering 300fs pulses and subsequently spectrally broadening and re-compressing the pulse using a hollow-core fiber and chirped mirrors. HR1 is available for user access, providing at this time  1.5mJ in 30 fs pulses at 1030nm central wavelength. HR2 was operated to deliver 10fs pulses with average power of 318W corresponding to a pulse energy of 3.18mJ.

A second 100KHz laser technology is employed in the mid-infrared (MIR) source, operated also at 100kHz but at 3100nm central wavelength \cite{Thire2018}, more suitable for a variety of applications including high order harmonics generation and attosecond pulses generation. The pulse energy is 70$\mu$J and the 20fs pulse duration corresponds to only two oscillation cycles of the electromagnetic field.

Single Cycle Laser group from ELI-ALPS worked on the development of few cycle laser pulses at 900nm central wavelength at 1KHz repetition rate \cite{Budriunas2017, Toth2020}. SYLOS2 laser delivers 4.8TW pulses of  more than 30mJ at a duration of  6.6 fs, corresponding to 2.3 optical cycles. The SYLOS3 upgrade which, is now under implementation in collaboration with Light Conversion company until end of 2022, will push the energy of the pulses to 120mJ coresponding to 15TW peak power. An additional alignment laser for SYLOS, SEA, can deliver similar pulses at 10 Hz repetition rate. The SEA laser produces 42.5mJ TW, 12fs pulses at 10Hz repetition rate. 

HF-2PW is a 2PW laser system planned to run at 10Hz. The key technology there is a pump laser able to deliver significant energy at 10Hz repetiton rate; in this case, a 50J@10Hz is under development for ELI-ALPS \cite{Falcoz2019}. Up to now, 10J at 10Hz repetition rate for 17fs pulses were declared operational, corresponding to 100W average power and more than 0.5 PW peak power. Complementary, a 100Hz laser system is planned to operate at up to 0.5J while the operational front-end delivers 8mJ in 25fs \cite{Cao2021}.  
\begin{figure}[tp]
	\centering
	\includegraphics[width=0.7\linewidth]{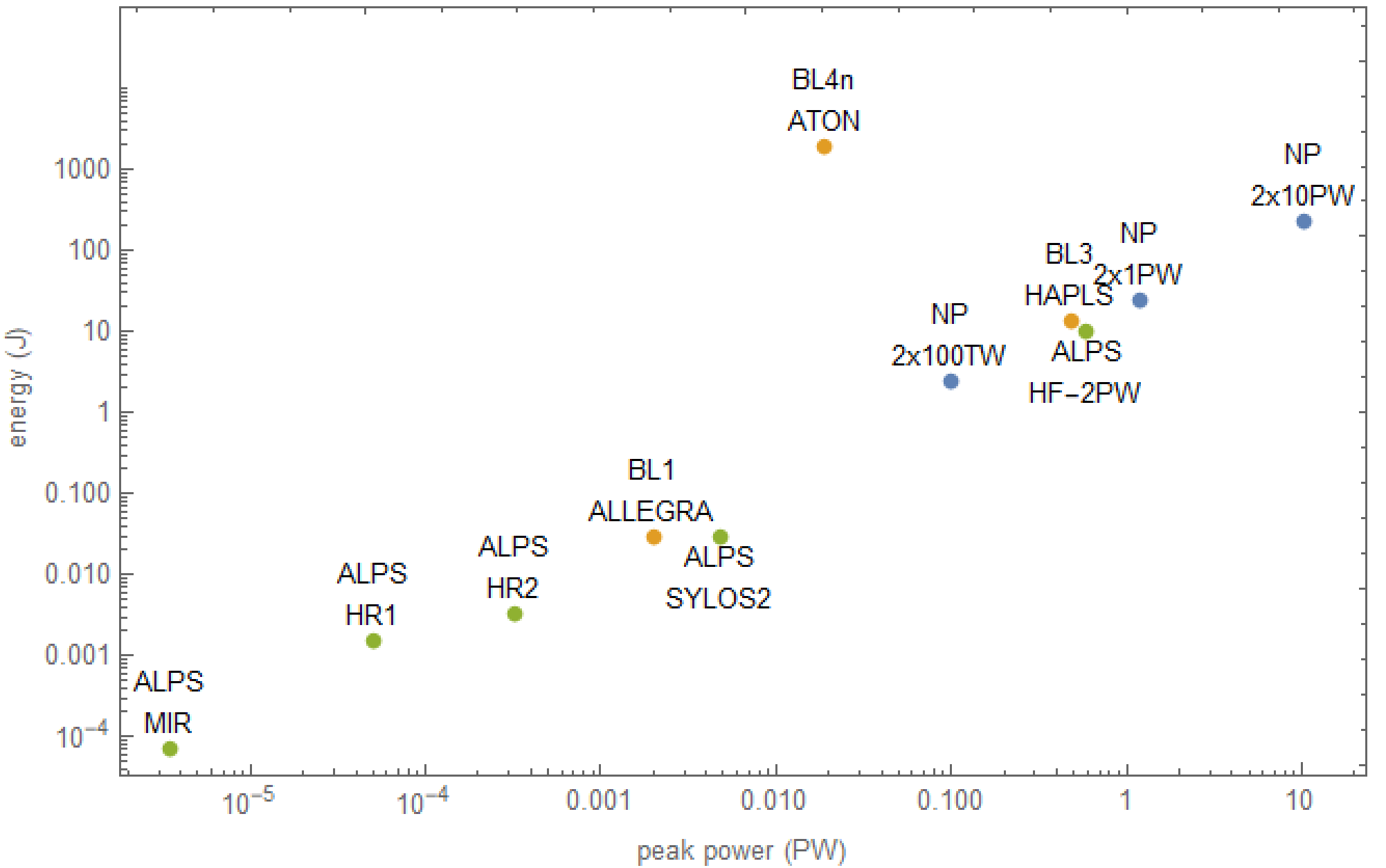}
	\caption{The reported tested peak power and energy parameters of the ELI main laser systems.}
	\label{fig:energy-power}
\end{figure}
\subsection{ELI-NP}
At ELI-NP, the High Power Laser System (HPLS) described in section \ref{sec:HPLS}, produced by a consortium within THALES company, is based on Ti:Sa oscillator amplified up to 10mJ at the repetition rate of 10Hz. Then the pulse is split in two and distributed for amplification in two arms, each arm containing Ti:sa crystals and pump lasers. Each arm has three outputs with dedicated optical compressor and diagnosis bench: at 100TW with 10Hz repetition rate, at 1PW with 1Hz repetition rate and at 10PW having one shot per minute repetition rate. The  pulse duration on each output is about 23fs. The laser was commissioned in 2019 \cite{Lureau2020}, experiments at 100TW output with nominal peak power were performed in the dedicated experimental area (E4) and in addition, demonstration of  10PW pulses generation was publicly performed twice to demonstrate the capability of 10PW pulses transport to the experimental areas E7 and E1-E6, in vacuum, at nominal parameters.
In addition to the HPLS, ELI-NP contracted a gamma source (VEGA) based on inverse Compton scattering of laser light with ultra-relativistic electrons obtained in a conventional accelerator to produce photons with energies per quanta up to 19MeV, that is under implementation and shall operate in 2023. 
\begin{figure}
	\centering
	\includegraphics[width=0.7\linewidth]{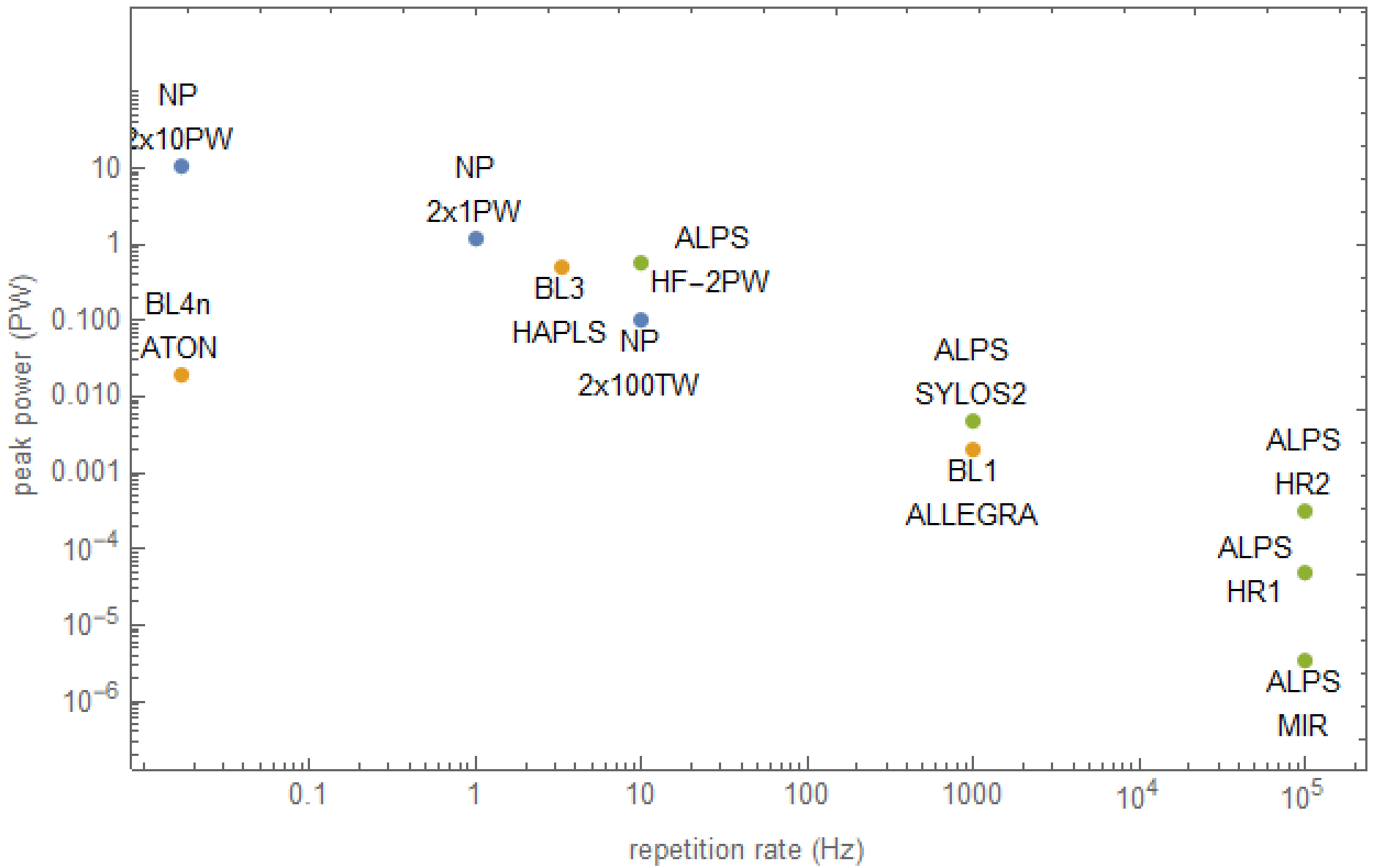}
	\caption{The achieved peak power as a function of the repetition rate for the ELI main laser systems.}
	\label{fig:peakpower-rep}
\end{figure}
\subsection{ELI status}
To summarize this section, figure \ref{fig:energy-power} illustrates the reported peak power and peak energy parameters of each major laser system in ELI facility, as reported in literature up to now.

Figure \ref{fig:energy-power} shows the high degree of complementarity of the ELI laser systems. The figure indicates that the  HPLS of ELI-NP reached the highest peak power at ELI, but also worldwide, meeting the specifications. The other laser systems belong also to best-in-class category, although they often did not reached the nominal design parameters, to date. Figure \ref{fig:peakpower-rep} presents the peak power as a function of the repetition rate of the laser system.

It is important to note that the figures \ref{fig:energy-power} and \ref{fig:peakpower-rep} have logarithmic scales on both axes. A simplified interpretation of the two figures would be that the increase of the repetition rate of the laser systems favors the industrial applications. For example, secondary sources of radiation can be produced with significant flux of particles. On the other hand, achieving higher peak power makes possible investigation of phenomena in new regimes, hence pushing the boundaries of our knowledge. Together, the industrial developments and the scientific research drive the progress of the human society.

\section{Summary}
The basic concept of peak intensity of ultrashort laser pulses was introduced. Its relation to the electric field, magnetic field and pressure were discussed, pointing out that the achievable values for these fields correspond to the largest ones ever produced by human kind in controlled manner. Then, the most powerful laser worldwide, HPLS from ELI-NP, was briefly described in order to accommodate the reader with specific CPA subsystems which are often used to reach extreme intensities.
 Finally, one of the most advanced research facilities worldwide, the Extreme Light Infrastructure, was summarized, in order to have an overview at the state of the art laser technology.
		
		\begin{acknowledgments}
			Extreme Light Infrastructure Nuclear Physics (ELI-NP) Phase II, is a project co-financed by the Romanian Government and the European Union through the European Regional Development Fund and the Competitiveness Operational Programme (1/07.07.2016, COP, ID 1334).  The contribution of the entire Thales and ELI-NP teams and collaborators are gratefully acknowledged. 
		\end{acknowledgments}


		

%

	\end{document}